\documentclass[conference]{IEEEtran}
\IEEEoverridecommandlockouts
\usepackage{cite}
\usepackage[numbers]{natbib}
\usepackage{amsmath,amssymb,amsfonts}
\usepackage{latexsym}
\usepackage{multirow}
\usepackage{algorithmic}
\usepackage{graphicx}
\usepackage{textcomp}
\usepackage{subfig}
\usepackage{xcolor}
\def\BibTeX{{\rm B\kern-.05em{\sc i\kern-.025em b}\kern-.08em
T\kern-.1667em\lower.7ex\hbox{E}\kern-.125emX}}

\makeatletter
\def\endthebibliography{%
\def\@noitemerr{\@latex@warning{Empty `thebibliography' environment}}%
\endlist
}
\makeatother
\begin{document}

\title{Heightmap Reconstruction of Macula on Color Fundus Images Using Conditional Generative Adversarial Networks}

\author{\IEEEauthorblockN{Peyman Tahghighi\IEEEauthorrefmark{1},
Reza A.Zoroofi\IEEEauthorrefmark{1},
Sare Safi \IEEEauthorrefmark{2}, and
Alireza Ramezani\IEEEauthorrefmark{3} \IEEEauthorrefmark{4}}
\IEEEauthorblockA{\IEEEauthorrefmark{1}School of Electrical and Computer Engineering\\
University of Tehran,
Tehran, Iran\\}
\IEEEauthorblockA{\IEEEauthorrefmark{2}Ophthalmic Research Center, Research Institute for Ophthalmology and Vision Science\\ Shahid Beheshti University of Medical Sciences, Tehran, Iran}
\IEEEauthorblockA{\IEEEauthorrefmark{3}Ophthalmic Epidemiology Research Center, Research Institute for Ophthalmology and Vision Science\\ Shahid Beheshti University of Medical Sciences, Tehran, Iran}
\IEEEauthorblockA{\IEEEauthorrefmark{4}Negah Specialty Ophthalmic Research Center\\ Shahid Beheshti University of Medical Sciences, Tehran, Iran
\\
Email: peyman.tahghighi@ut.ac.ir}}

\maketitle

\begin{abstract}
For screening, 3D shape of the eye retina often provides structural information and can assist ophthalmologists to diagnose diseases. However, fundus images which are one the most common screening modalities for retina diagnosis lack this information due to their 2D nature. Hence, in this work, we try to infer about this 3D information or more specifically its heights. Recent approaches have used shading information for reconstructing the heights but their output is not accurate since the utilized information is not sufficient. Additionally, other methods were dependent on the availability of more than one image of the eye which is not available in practice. In this paper, motivated by the success of Conditional Generative Adversarial Networks(cGANs) and deeply supervised networks, we propose a novel architecture for the generator which enhances the details in a sequence of steps. Comparisons on our dataset illustrate that the proposed method outperforms all of the state-of-the-art methods in image translation and medical image translation on this particular task. Additionally, clinical studies also indicate that the proposed method can provide additional information for ophthalmologists for diagnosis.
\end{abstract}


\section{Introduction}
\par
A major and ubiquitous imaging technique for the eye is the fundus camera. However, due to its 2D nature, it lacks a considerable amount of data such as information about elevations in different parts and also the different number of views. Optical Coherence Tomography (OCT) \cite{29} is an expensive but vital tool for retinal structure diagnosis which provides ophthalmologists with valuable information, enabling them to diagnose most of the retinal diseases such as glaucoma, diabetic and
macular degeneration\cite{30}. Nevertheless, owing to the cost of using this system, it is not common and using fundus images is mostly common for screening.
\par

\begin{figure}[h]
\centering
\includegraphics[clip,trim=5.5cm 19.0cm 3cm 1cm width=1.0\linewidth,scale=0.8]{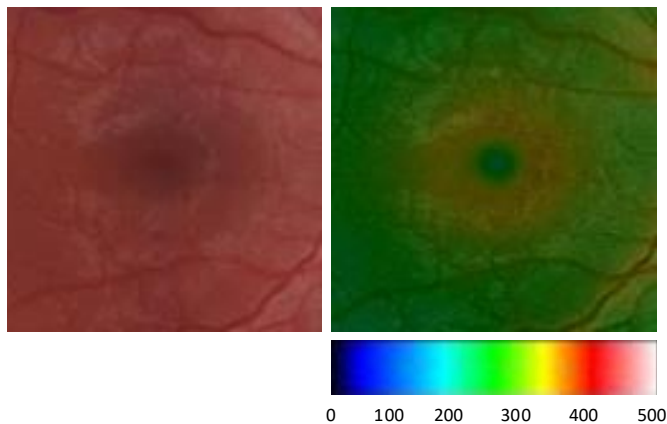}
\caption{The left and right image represent the correspondence between a fundus image and its thickness map image. As can be seen, each pixel's color value of the image on the right indicates a height according to color gradients below which ranges from 0$\mu$m to 500$\mu$m.}
\label{fig1}
\end{figure}
Shape from shading (SFS) \cite{49} is the only method applied to this problem for the reconstruction of the height of a single fundus image \cite{4}. However, the result generated by this method is unreal and exaggerated \cite{3} which limits its usage in real diagnosis. Furthermore, disparity map estimation, which is one of the common methods for 3D reconstruction was also applied to this problem \cite{41}, but, it totally depends on the availability of two stereo images from both eyes which is not practical. Hence, devising a method to automatically generate a heightmap image from a given fundus image is crucial.

\par

In recent years, with the advent of Conditional Generative Adversarial Networks (cGANs) \cite{24,31}, many researchers used this methods for image transformation tasks \cite{24,39,40,76,77}. Recently cGANs have drawn attention in the medical field especially in the
task of image-to-image translation such as translating between CT and PET images \cite{74,25}, denoise and correct the motion in medical images such as denoising CT images \cite{33} and segmenting medical images \cite{34,78,79}. Most of these methods benefited from the U-Net architecture \cite{37} and its extension U-Net++ \cite{80} which first were proposed for image segmentation.
\par

\begin{figure*}[h]
\centering
\captionsetup[subfigure]{justification=justified,singlelinecheck=false}
\includegraphics[clip,trim=0.06cm 8cm 35.0cm 0.1cm, scale=0.6]{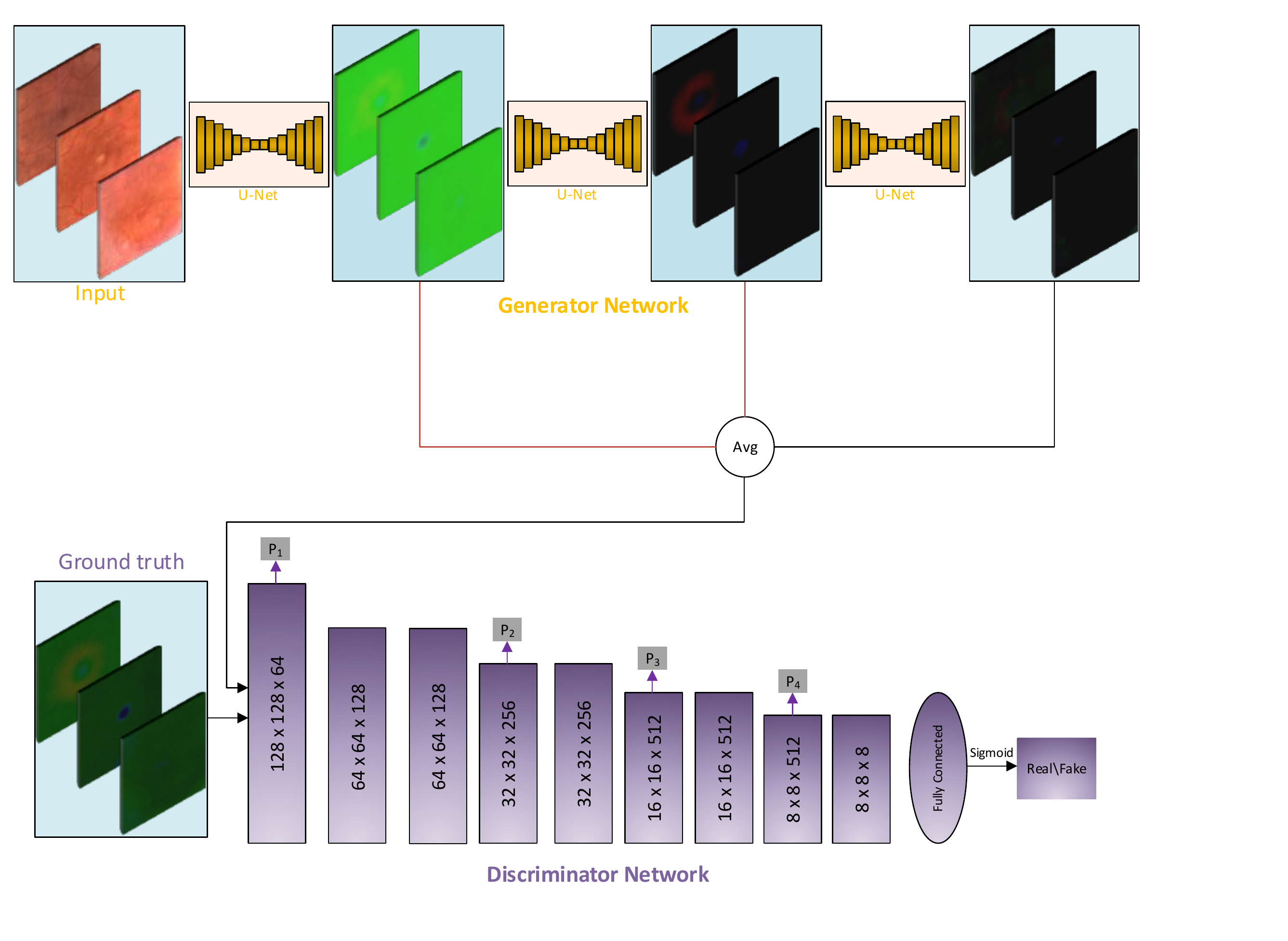}
\caption{The architecture of generator and discriminator in our proposed method. The red arrows indicate deep supervision.}
\label{fig2}
\end{figure*}

Considering Figure \ref{fig1}, since our problem can be seen as an image translation task in which we want to predict a color image in which each of the pixels represents a height from 0 $\mu m$ to 500 $\mu m$ from a fundus image targeted on the macula area, cGANs can be utilized in this problem. In fact, height information is one of the crucial information that OCT devices provide to ophthalmologists and is missing in color fundus images due to their 2D nature. Hence, by extracting such information from only a fundus image, we can ease the diagnosis and management of retinal diseases with macular thickness changes.

\par

In our cGAN setting, we used a stack of three U-Nets for generator network which we averaged on the output of them for deep supervision. Furthermore, in order to avoid problems of traditional GANs, we used Least Squares Adversarial Loss \cite{65} instead along with perceptual loss \cite{39} and L2-loss. For the discriminator network, we used an image-level discriminator that classifies the whole image as real or fake. To the best of our knowledge, this is the first research paper on predicting the heightmap of the macula area on fundus images using Deep Neural Networks (DNNs). We compared our approach qualitatively and quantitatively on our dataset with state-of-the-art methods in image translation and medical image translation. Furthermore, we studied the application of our method on real diagnosis cases which showed that reconstructed heightmaps can provide additional information to ophthalmologists.

\section{Methods}
\subsection{Network architecture}
Our proposed architecture is depicted in Figure \ref{fig2}. As can be seen, our proposed generator architecture is consists of three stacked U-Nets. Additionally, since discriminative features in deep layers will contribute to higher performance \cite{69}, we used the output of the first two U-Net layers for deep supervision. By doing so, our network tries to learn a meaningful representation for these deep layers which will directly contribute to the final outcome. Moreover, since the amount of detail is of significance in this work, we decided to use the average operator instead of the max operator for the final output \cite{80}. Another advantage of this architecture is that by using a stack of U-Nets, each layer can add its own level of detail to the final outcome and the details can gradually refine. Furthermore, as can be seen, we used an image-level discriminator that classifies the whole image as real or fake which using a series of Convolution-BatchNorm-LeakyReLU layers. Finally, we used $1^{th},4^{th},6^{th} $ and $8^{th}$ layer of the discriminator network to compute perceptual loss \cite{40} between generated image and ground-truth image as a supervisory signal with the aim of better output.

\subsection{Objective functions}
Our final loss function is composed of three parts which will be discussed in this section.
\subsubsection{Least-squares adversarial loss}
cGANs are generative models that learn mapping from observed image $x$ and random noise vector $z$ to $\hat{y}$, using generator $G: {x,z} \rightarrow \hat{y}$. Then the discrimnator $D$ aims to classify the concatenation of the source image $x$ and its corresponding ground-truth image $y$ as real $D(x,y) = 1$, while classifying $x$ and the transformed image $\hat{y}$ as fake, $D(x,\hat{y}) = 0$.
\par
GANs suffer from different problems such as mode collapse or unstable training procedure despite producing plausible results in many tasks \cite{24}. Hence, to avoid such problems we used Least Square Generative Adversarial Networks (LSGANs) \cite{65}. The idea of LSGAN is that even samples which were classified correctly can provide signals for training. Hence, for achieving this aim, we adopted the least-squares loss function instead of the traditional cross-entropy loss used in normal cGAN to penalize data on the right side of the decision boundary but very far from it. Using this simple idea we can provide gradients even for samples that are correctly classified by the discriminator. The loss function of LSGAN for both discriminator and generator can be written as follows:

\begin{flalign}
&\underset{D}{\min}\ L_{LSGAN}(D) = \frac{1}{2} \mathbb{E}_{x,y} \Big[\big(D(x,y) - b\big) ^2\Big] \ + \nonumber \\
&\qquad \qquad \qquad \nonumber \frac{1}{2} \mathbb{E}_{x,z}\Big[\Big(D\big(x,G(x,z) \big)-a\Big)^2\Big] \nonumber \\
&\underset{G}{\min}\ L_{LSGAN}(G) = \frac{1}{2} \mathbb{E}_{x,z}\Big[\Big(D\big(x,G(x,z) \big)-c\Big)^2\Big]&&
\end{flalign}

This loss functions directly operates on the logits of the output, where $a = 0$ and $c = b= 1$.

\subsubsection{Pixel reconstruction loss}
Relying solely on the adversarial loss function do not produce consistent results \cite{36}. Therefore, we also used pixel reconstruction loss here but we opted for L2-loss rather than widely used L1-loss since it performed better in reconstructing details in this specific task. The equation for L2-loss is as below:
\begin{equation}
\begin{split}
L_{L2}(G) = \mathbb{E}_{x,y,z}\big[\parallel y-G(x,z)\parallel^{2}_{2}\big]
\label{eq3}
\end{split}
\end{equation}
\subsubsection{Perceptual loss}
Despite producing plausible results using only two aforementioned loss functions, since the generated image is blurry \cite{25} and especially in medical diagnosis small details are of significance, we used perceptual loss \cite{28} to improve the final result. As a matter of fact, using only L2-Loss or L1-Loss results in outputs that maintain the global structure but it shows blurriness and distortions \cite{40}. Furthermore, per-pixel losses fail to capture perceptual differences between input and ground-truth images. For instance, when we consider two identical images only shifted by some small offset from each other, per-pixel loss value may vary considerably between these two, even though they are quite similar \cite{39}. However, by using high-level features extracted from layers of a discriminator, we can capture those discrepancies and can measure image similarity more robustly \cite{39}. In our work, since discriminator network also has this capability of perceiving the content of images and difference between them and pre-trained networks on other tasks may perform poorly on other unrelated tasks, we used hidden layers of discriminator network \cite{25,28} to extract features as illustrated in the second row of the Figure \ref{fig2}. The mean absolute error for $i^{th}$ hidden layer between the generated image and the ground-truth image is then calculated as :
\begin{equation}
\begin{split}
P_i\big(G(x,z),y\big) = \frac{1}{w_i h_i d_i} \parallel D_i\big(x,y\big) - D_i\big(x,G(x,z)\big) \parallel _{1}
\label{eq4}
\end{split}
\end{equation}
which $w_i$,$h_i$ and $d_i$ denote width, height and depth of the $i^{th}$ hidden layer respectively and $D_i$ means the output of $i^{th}$ layer of the discriminator network. Finally, perceptual loss can be formulated as :
\begin{equation}
\begin{split}
L_{perceptual} = \sum_{i=0}^{L}\lambda_i P_i\big(G(x,z),y\big)
\label{eq5}
\end{split}
\end{equation}
Where $\lambda_i$ in equation \ref{eq5} tunes the contribution of $i^{th}$ utilized hidden layer on the final loss.
\par
Finally, our complete loss function for the generator is as below:
\begin{equation}
\begin{split}
L = \alpha_{1}L_{perceptual} + \alpha_{2}L_{L2} + \alpha_{3}L_{LSGAN}
\label{eq6}
\end{split}
\end{equation}
where $\alpha_1$, $\alpha_2$ and $\alpha_3$ are the hyperparameters that balance the contribution of each of the different losses.

\begin{table*}[!ht]
\centering
\begin{tabular}{l|c|c|c|c}
Method & \multicolumn{1}{l|}{SSIM} & \multicolumn{1}{l|}{PSNR(dB)} & \multicolumn{1}{l|}{LPIPS} & \multicolumn{1}{l}{MSE} \\ \hline
pix2pix \cite{27} & 0.8596 & 33.9523 & 2.25e-03 & 0.0068 \\ \cline{1-1}
PAN \cite{28} & 0.8612 & 33.8512 & 2.37e-04 & 0.0053 \\ \cline{1-1}
MedGAN \cite{25} & 0.8659 & 33.2958 & 5.61e-05 & 0.0048 \\ \cline{1-1}
Proposed Method & \textbf{0.8823} &\textbf{ 34.6733} & \textbf{1.81e-05} &\textbf{0.0033}
\end{tabular}
\caption{Quantitative comparison between proposed method and other methods.}
\label{tb1}
\end{table*}

\begin{figure*}[h]
\centering

\subfloat{
\begin{minipage}{
0.19\textwidth}
\includegraphics[width=1\textwidth,height = 1\textwidth]{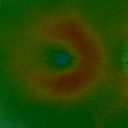}
\end{minipage}}\hspace{0.25cm}
\subfloat{\begin{minipage}{
0.19\textwidth}
\includegraphics[width=1\textwidth,height = 1\textwidth]{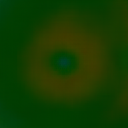}
\end{minipage}}\
\subfloat{\begin{minipage}{
0.19\textwidth}
\includegraphics[width=1\textwidth,height = 1\textwidth]{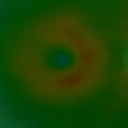}
\end{minipage}}\
\subfloat{\begin{minipage}{
0.19\textwidth}
\includegraphics[width=1\textwidth,height = 1\textwidth]{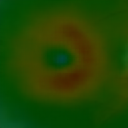}
\end{minipage}}\
\subfloat{\begin{minipage}{
0.19\textwidth}
\includegraphics[width=1\textwidth,height = 1\textwidth]{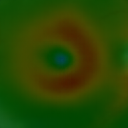}
\end{minipage}}
\vfill

\setcounter{subfigure}{0}%

\subfloat[Ground-truth]{
\begin{minipage}{
0.19\textwidth}
\includegraphics[width=1\textwidth,height = 1\textwidth]{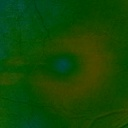}
\end{minipage}}\hspace{0.25cm}
\subfloat[pix2pix]{\begin{minipage}{
0.19\textwidth}
\includegraphics[width=1\textwidth,height = 1\textwidth]{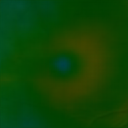}
\end{minipage}}\
\subfloat[PAN]{\begin{minipage}{
0.19\textwidth}
\includegraphics[width=1\textwidth,height = 1\textwidth]{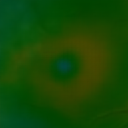}
\end{minipage}}\
\subfloat[MedGAN]{\begin{minipage}{
0.19\textwidth}
\includegraphics[width=1\textwidth,height = 1\textwidth]{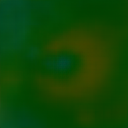}
\end{minipage}}\
\subfloat[Proposed method]{\begin{minipage}{
0.19\textwidth}
\includegraphics[width=1\textwidth,height = 1\textwidth]{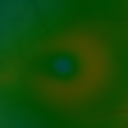}
\end{minipage}}
\setcounter{subfigure}{0}%

\caption{Qualitative comparison between the proposed method and other methods.}
\label{fig3}

\end{figure*}

\section{Experiments}
\subsection{Dataset and setup}
The dataset was gathered from TopCon DRI OCT Triton captured at Negah Eye Hospital. Our dataset contains 3407 color fundus-heightmap pair images. Since the images in our dataset were insufficient, we used data augmentation for better generalization. However, flipping images were the only available option since by rotating images for example by 90$^\circ$,  we have vessels in a vertical position which is impossible in fundus imaging or changing brightness will change the standard brightness of a fundus image. Hence, we flipped every image in 3 different ways to generate 4 samples. Consequently, we had 13,628 images which we used 80$\%$ for training, 10$\%$ for validation and 10$\%$ for testing.
\par
We used Tensorflow 2.0 \cite{43} for implementing our network. We also used Adam optimizer \cite{44} with an initial learning rate of $1e^{-3}$ with a step decay of $0.9$ per $30$ steps. Moreover, we used the batch size of 8 and trained for 250 epochs to converge. Additionally, we set $\lambda_1= 5.0$, $\lambda_2 = 1.0$, $\lambda_3 = 5.0$ and $\lambda_4 = 5.0$ in Equation \ref{eq5} by trial-and-error and considering the contribution of each of them as discussed in \cite{28,39}.

\subsection{Evaluation metrics}
we utilized a variety of different metrics to evaluate our final outcomes quantitatively such as Structural Similarity Index (SSIM) \cite{45}, Mean Squared Error (MSE) and Peak Signal to Noise Ratio (PSNR). Nevertheless, these measures are insufficient for assessing structured outputs such as images, as they assume pixel-wise independence \cite{46}. Hence, we used Learned Perceptual Image Patch Similarity (LPIPS) \cite{46} which can outperform other measures in terms of comparing the perceptual quality of images. For this measure, we used features extracted from the $1^{th},4^{th},6^{th} $ and $8^{th}$ layer of the discriminator network for both generate heighmap and ground-truth.

\subsection{Comparison with other techniques}
\label{sec5.8}
Since this is the first method for the reconstruction of the heightmap of color fundus images using DNNs, there were no other method to directly compare our proposed method with. Hence, we compared our results with common methods which utilized cGANs such as pix2pix \cite{27}, PAN \cite{28} and MedGAN \cite{25}. As can be seen in Table \ref{tb1}, Pix2pix achieved the worst results since it does not use deep supervision and it is based on L1-loss. PAN did slightly better in SSIM and MSE metrics. However, there is a huge difference between PAN and pix2pix in terms of LPIPS since PAN uses perceptual loss in the training procedure which proves the impact of using perceptual loss for training cGANs. MedGAN was designed especially for medical image translation and, as a result, It performs better in comparison to previous general methods. However, the results are inferior to our proposed method. In fact, since we are using deep supervision in this method and carefully tuned the parameters for this particular problem, we achieve a higher value in all metrics.

\par

Considering the qualitative comparison in Figure \ref{fig3}, our proposed method outperformed others in terms of reconstruction of the details. As can be seen, pix2pix missed some of the important details in first example such as bright red spots. PAN performed better at reconstructing the highly elevated parts in the first row, but it failed to reconstruct the correct shape for the second example. Finally, since MedGAN is specially designed for medical tasks, it outperformed the aforementioned methods in terms of output quality, but it was outperformed by our method and the proposed method generated the best quality images.

\begin{figure*}[h]
\captionsetup[subfigure]{labelformat=empty,position=top}
\centering

\subfloat[Fundus]{
\begin{minipage}{
0.15\textwidth}
\includegraphics[width=1\textwidth,height = 1\textwidth]{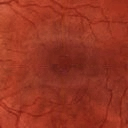}
\end{minipage}}\
\subfloat[Proposed method]{\begin{minipage}{
0.15\textwidth}
\includegraphics[width=1\textwidth,height = 1\textwidth]{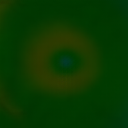}
\end{minipage}}\
\subfloat[Ground-truth]{\begin{minipage}{
0.15\textwidth}
\includegraphics[width=1\textwidth,height = 1\textwidth]{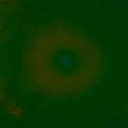}
\end{minipage}}\hspace{0.25cm}
\subfloat[Fundus]{\begin{minipage}{
0.15\textwidth}
\includegraphics[width=1\textwidth,height = 1\textwidth]{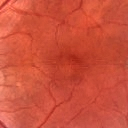}
\end{minipage}}\
\subfloat[Proposed method]{\begin{minipage}{
0.15\textwidth}
\includegraphics[width=1\textwidth,height = 1\textwidth]{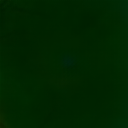}
\end{minipage}}\
\subfloat[Ground-truth]{\begin{minipage}{
0.15\textwidth}
\includegraphics[width=1\textwidth,height = 1\textwidth]{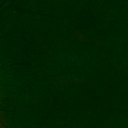}
\end{minipage}}

\captionsetup[subfigure]{labelformat=parens,position=bottom}
\setcounter{subfigure}{0}%

\caption{Some examples from images classified as positive by two ophthalmologists.}
\label{fig4}

\end{figure*}

\subsection{Perceptual studies}
\label{sectionPerceptualStudies}
To judge usefulness in real diagnosis cases, we conducted an experiment in which two experienced ophthalmologists were presented a series of trails each containing a reconstructed heightmap, fundus image and the ground-truth heightmap from the test set to investigate if the reconstructed heightmap and fundus image pair gives more information for the diagnosis of any retinal disease in comparison to the situation in which we have the fundus image only. For this experiment, two ophthalmologists rated each image from zero to three which zero means that the reconstructed heightmap does not add more information(negative). From one to three, the heightmap is in positive class and the rating represents the level of information that it provides.
\par

As can be seen in Table \ref{tb2}, ophthalmologist 1 classified all images as useful for diagnosis and the mean score for all of the images is 1.94. Additionally, ophthalmologist 2 classified 92$\%$ of the outputs as positive and the mean square is 1.84. This experiment shows that although the reconstructed heightmap may seem more blurry than the ground-truth, it can still provide additional information for diagnosis especially about height information in different regions. Some positive examples are illustrated in Figure \ref{fig4}.

\begin{table}[h]
\centering
\begin{tabular}{l|c|c|c}
\multirow{2}{*}{} & \multicolumn{2}{c|}{Score} & Classification \\ \cline{2-4}
& Mean & SD & Positive \% \\ \hline
Ophthalmologist 1 & 1.94 & 0.7669 & 100.00 \\ \cline{1-1}
Ophthalmologist 2 & 1.84 & 0.9553 & 92.00
\end{tabular}
\caption{Results of perceptual study.}
\label{tb2}
\end{table}

\section{Conclusion and Discussion}
In this paper, we proposed a novel framework to automatically generate heightmap images from the macula on a color fundus image. Leveraging the power of cGANs and deep supervision, this new paradigm can generate heightmaps which can provide additional information for ophthalmologists for diagnosis. Furthermore, comparisons indicate our method outperformed other methods in the task of image and medical image translation in terms of SSIM, PSNR, MSE and LPIPS metrics. However, this work is not free from limitations and further improvement is essential to improve the quality of the outputs. By investigation, we found that in most failed cases, the poor image quality caused the system to fail especially in cases in which the fundus image is blurred. This suggests that in future works, a proper pre-processing step should be employed to de-blur fundus images and improve their quality before feeding them into the network.
\par
Finally, considering the perceptual studies which show that our reconstructed heightmap can provide additional information for the diagnosis, in future researches, we can utilize the generated heightmaps to train a classifier that is able to automatically diagnose diseases that were impossible to diagnose automatically before.

\section{Acknowledgment}
The authers are grateful to Dr.Ahmadieh (Ophthalmologist) for grading and classifying images for our experiment.

\bibliographystyle{unsrt}
\bibliography{Refrences}

\end{document}